# THE APPLICATION OF SPECIAL MATRIX PRODUCT TO DIFFERENTIAL QUADRAUTRE SOLUTION OF GEOMETRICALLY NONLINEAR BENDING OF ORTHOTROPIC RECTANGULAR PLATES

W. Chen (Corresponding author) and

Permanent mail address: P. O. Box 2-19-201, Jiangshu University of Science & Technology, Zhenjiang City, Jiangsu Province 212013, P. R. China

Present mail address (as a JSPS Postdoctoral Research Fellow): Apt.4, West $1^{st}$ floor, Himawari-so, 316-2, Wakasato-kitaichi, Nagano-city, Nagano-ken, 380-0926, JAPAN

E-mail: chenw@homer.shinshu-u.ac.jp

Permanent email: chenwwhy@hotmail.com

C. Shu

Department of Mechanical and Production Engineering, National University of Singapore, Singapore

W. He

Department of Electrical Engineering, Jiangsu University of Science & Technology, Zhenjiang, Jiangsu 212013, P. R. China.

## ABSTRACT

The Hadamard and SJT product of matrices are two types of special matrix product. The latter was first defined by Chen [1]. In this study, they are applied to the differential quadrature (DQ) solution of geometrically nonlinear bending of isotropic and orthotropic rectangular plates. By using the Hadamard product, the nonlinear formulations are greatly simplified, while the SJT product approach minimizes the effort to evaluate the Jacobian derivative matrix in the Newton-Raphson method for solving the resultant nonlinear formulations. In addition, the coupled nonlinear formulations for the present problems can easily be decoupled by means of the Hadamard and SJT product. Therefore, the size of the simultaneous nonlinear algebraic equations is reduced by two-thirds and the computing effort and storage requirements are alleviated greatly. Two recent approaches applying the multiple boundary conditions are employed in the present DQ nonlinear computations. The solution accuracies are improved obviously in comparison to the previously given by Bert et al. [9]. The numerical results and detailed solution procedures are provided to demonstrate the superb efficiency, accuracy and simplicity of the new approaches in applying DQ method for nonlinear computations.

## 1. INTRODUCTION

The differential quadrature (DQ) method was introduced by Bellman and his associates [1, 2]. Since then, the method has been applied successfully to a broad range of problems [2-16]. A striking merit of the method is of high efficiency in computing nonlinear problems [2, 3, 5, 7-11, 14-16]. Compared with the standard numerical techniques such as the finite element and finite difference methods, the DQ method produces solution of reasonable accuracy with relatively small computational effort. This method also does not require seeking the trial functions satisfying boundary conditions as in the Rayleigh-Ritz and Galerkin methods [12]. Therefore, it is easily used for practical purposes. It has been found that the DQ method is closely related with the collocation (or pseudo-spectral) methods [12, 17, 18]. The principal advantages of the DQ method over the latter are its simplicity and ease in using grid spacing without restriction [8, 18].

The geometrically nonlinear behavior of thin plates is usually described by the von Karman equations and has become a benchmark problem for testing numerical solutions to nonlinear partial differential equations. Bert et al. [9] employed the DQ method to solve the static von Karman equations in analyzing geometrically nonlinear bending of isotropic and orthotropic rectangular plates. The work shows that the DQ method is an efficient numerical technique in solving complex nonlinear problems by comparing with the finite element and finite difference methods. However, the application is not very successful in plate with all edge simply supported. The solution procedure is also much more complex in comparison to the solution of linear problems. Nonlinear problems presently are considered to be in the territory of high performance workstation and supercomputer. With the development of personal computers, it has become possible to develop a computationally inexpensive approach for the nonlinear computations via PC computer. The DQ method should be of choice in this respect. However, in solving complex multidimensional nonlinear problems, for example, von Karman equations of plates, computational effort and storage requirements are still rather high even in the DQ method. Therefore, if the DQ method attempts to undertake more nonlinear computations via PC computer, its computing cost need obviously be reduced. Recently, some remarkable advances are achieved to simplify application and improve efficiency and accuracy of the DQ method. In this study, values of some new techniques in the DQ computing are verified through the solution of the nonlinear geometrical bending of plates. Especially, the work validates the applicability, simplicity and high efficiency of special product approach in the DQ nonlinear computations. It should be pointed out that due to a great reduction in the computing effort and storage requirements by using the new techniques, all results presented here were accomplished on a 386 personal computer with only 4MB memory. In contrast, Bert et al. [9] used an IBM framework computer for the same task. In the following, innovations in this work are briefly introduced.

First, the Hadamard product of matrices and the DQ matrix approximate formulas are used to express the formulation of nonlinear partial differential operator in explicit and easily programmable matrix form. The DQ analog formulas in matrix form [20] can be viewed as simple and compact version of the traditional polynomial approximation of multidimensional problems [4]. By using these formulas, the formulation effort can be simplified. The often-used ordinary matrix product rises from the concept of linear transformation and is extended to handle the nonlinear problems. However, since nonlinear problems are actually different from linear ones, the ordinary matrix product seems not to undertake the task of nonlinear analysis and computations very well. The Hadamard product is a kind of special matrix product and not well known to the numerical computation community. It was found that the Hadamard product provides an explicit, compact and convenient approach to formulate the nonlinear differential operators in the DQ method [15] as well as the other numerical techniques [16]. Second, the SJT product was first introduced by Chen and Zhong [15] and Chen [16] as an efficient and simple algorithm to compute the exact Jacobian derivative matrix in the Newton-Raphson solution of the nonlinear formulations in the Hadamard product form. The SJT product is also a kind of special matrix product. It is emphasized that the SJT product approach may require minimal computing effort in all possible approaches for the same task. More importantly, the Hadamard and SJT product provides a decoupling technique in solving the coupled nonlinear DQ analog equations of the von Karman equations. By comparing with Bert et al. [9], the resulting size of the nonlinear simultaneous algebraic equations is reduced by two-thirds. Therefore, the computational effort and storage requirements are alleviated significantly. Third, a recent approach applying multiple boundary conditions in the DQ analysis of high order boundary value problems, proposed by Wang and Bert [21], is employed for the solution of the geometrically nonlinear simply-supported plates. It was verified that in the linear cases, the accuracies of the DQ solutions using Wang and Bert's approach were evidently improved in comparison to using the conventional so-called $\delta$ approach [12, 21, 22]. Wang and Bert's approach is also effective for the other boundary conditions except for the clamped-clamped (C-C) boundary condition [21, 22] and free edge of plate. Chen and Yu [23] proposed another different approach to improve accuracy and eliminate instability caused by $\delta$ effect in the conventional $\delta$ approach. Chen and Yu's approach is applicable for problems with any boundary constraint. In this study, the approach is applied to analyze the plate of clamped edges. An improvement on this approach is also presented to avoid extra four equations of corner points in the resulting formulation and reduce the size of the resultant algebraic systems. It is believed that these two new approaches applying multiple boundary conditions are first used in the DQ solution of nonlinear boundary value problems. To avoid confusion, all three approaches are denoted as the DQδ approach [6, 7, 9, 10], DQWB approach by Wang and Bert [21], and DQCY approach by Chen and Yu [23]. Fourth, the Chebyshev grid points are used in the present numerical study. As is expected, the DQ method using such grid points has faster rate of convergence than using the equally spaced grid points in [9]. Finally, some conclusions are drawn based on the results reported herein.

## 2. APPROACHES APPLYING BOUNDARY CONDITIONS AND MATRIX APPROXIMATE FORMULAS IN THE DQ METHOD

The differential quadrature method approximates the partial derivative of a function at a given discrete point as a weighted linear sum of function values at all of the discrete points along the respective variable direction in the entire domain of the variable. The m-th order derivative of a single function f(x) at a given discrete grid point i can be approximated by the DQ method with N discrete grid points [2-6], namely,

$$\left.\frac{\partial^m f}{\partial x^m}\right|_{x=x_i} = \sum_{j=1}^{N} w_{ij}^{(m)} f_j \quad i=1,2,\ldots,N. \tag{2-1}$$

where $f_j = f(x_j)$, $w_{ij}^{(m)}$ are the corresponding DQ weighting coefficients and can be determined by requiring that Eq. (2-1) be exact for all polynomials less than or equal to N-1. The explicit formulas [11, 17] were developed to conveniently obtain the accurate DQ weighting coefficients. For the weighting coefficients of the 1st order derivatives, we have

$$w_{ij}^{(1)} = \frac{1}{x_j - x_i} \prod_{\substack{k \neq i \\ k \neq j \\ i \neq j}}^{N} \frac{x_i - x_k}{x_j - x_k}, \qquad i=1,2,\ldots N \text{ and } j = 1,2,\ldots,N. \qquad (2\text{-}2a)$$

$$w_{ii}^{(1)} = -\sum_{k \neq i}^{N} \frac{1}{x_i - x_k}, \qquad i=1,2,\ldots N \qquad (2\text{-}2b)$$

The weighting coefficients for high order derivatives can be generated by recursion formulas [17]:

$$w_{ij}^{(m+1)} = m\left( w_{ij}^{(1)} w_{ii}^{(m)} - \frac{w_{ij}^{(m)}}{x_i - x_j} \right), \qquad i \neq j \qquad (2\text{-}3a)$$

$$w_{ii}^{(m+1)} = -\sum_{j \neq i}^{N} w_{ij}^{(m+1)}, \qquad (2\text{-}3b)$$

where the superscript (m) and (m+1) denote the order of the derivative.

In the present study, besides the uniform points, the zeros of Chebyshev polynomials of the first kind are considered, namely,

$$r_i = \cos\frac{(2i-1)\pi}{2N}, \qquad -1<r_i<1, \quad i=1, 2, \ldots, N. \qquad (2\text{-}4)$$

where $r_i$ is the i-th root of the N order Chebyshev polynomials. The zeros do not included the end points of the normalized domains ($0 \leq x \leq 1$ or $-1 \leq x \leq 1$), while in the present case, the end points ($x=0,1$ or $x=\pm 1$) are needed to force boundary conditions. Therefore, an algebraic transformation is required, namely,

$$x_i = \frac{r_i - 1}{r_N - r_1}, \qquad 0 \leq x_i \leq 1 \qquad i=1, 2, \ldots, N. \qquad (2\text{-}5a)$$

$$y_j = \frac{r_j - 1}{r_M - r_1}, \qquad 0 \leq y_j \leq 1 \qquad j=1, 2, \ldots, M. \qquad (2\text{-}5b)$$

The above formulas produce the coordinates of sampling points. With such grid points, a very simple formula can be obtained for computing the DQ weighting coefficients of the 1st order derivatives.

$$w_{ij}^{(1)} = \frac{(-1)^{(i-j)}(r_N - r_1)}{r_i - r_j}\sqrt{\frac{(1-r_j^2)}{(1-r_i^2)}}, \qquad i \neq j \qquad (2\text{-}6a)$$

$$w_{ii}^{(1)} = \frac{1}{2}\frac{r_i(r_N - r_1)}{1 - r_i^2}, \qquad i=1,2,\ldots, N \qquad (2\text{-}6b)$$

The weighting coefficients for high order derivatives can be easily computed by using recurrence formulas (2-3a, b). More details about the DQ method see references [2, 4, 6, 11, 17]. It should be especially noted that in order to differ the variants from the DQ weighting coefficients more clearly, the notations for the DQ weighting coefficients of the 1$^{st}$, 2$^{nd}$, 3$^{rd}$ and 4$^{th}$ order derivatives are represented as $A_{ij}$, $B_{ij}$, $C_{ij}$ and $D_{ij}$, respectively, in all subsequent sections. $\overline{A}$, $\overline{B}$, $\overline{C}$ and $\overline{D}$ represent the corresponding modified coefficient matrices with built-in boundary conditions by the DQWB or DQCY approach.

## 2.1 Three approaches applying multiple boundary conditions

The governing equations in the structural mechanics usually involve the fourth-order derivatives, and the two conditions need be satisfied at each boundary. Therefore, the problems are more complex in comparison to the differential systems of no more than second order such as the Poisson equation and convection-diffusion equation, which do not involve more than one boundary condition at each boundary end of physical domain. Some careful considerations are necessary how to properly implement the double boundary conditions at each edge. The problem is actually the DQ solution of high-order boundary value problems with multiple boundary conditions. In what follows we discuss three existing approaches for such problems.

The earliest DQδ approach, proposed by Bert et al. [6] and widely used in literature [7, 9, 10], enforces the geometry boundary condition at boundary points and derivative boundary conditions at the δ points, which have very small distance δ ($\delta \cong 10^{-5}$ in dimensionless value [22]) away from the boundary. Thus, the approach can not satisfy derivative boundary conditions exactly at boundary points and the accuracies of the solutions are affected. The solution accuracy depends on the proper choice of δ. If the value of δ is small enough, the approach produce good results in some situations such as clamped condition, however, failed to work well in the other situations such as simply-supported and free edges [10, 21, 22]. On the other hand, too small δ will deteriorate the computations. δ is usually determined by trial and error for different cases [22], which is a rather tedious work. Arbitrariness in the choice of the δ value may introduce the unexpected oscillation behavior of the solutions. In addition, the number of grid points in the DQδ approach can not be large due to the ill-conditioned matrices caused by the δ grid spacing.

To overcome the drawbacks in the DQδ approach, some new techniques were presented in the application of the multiple boundary conditions. Wang and Bert [21] developed a DQWB approach by incorporating the boundary conditions into the DQ weighting coefficient matrices in advance, and then the weighting coefficients with built-in boundary conditions are employed to analogize the governing equations of problems of interests. The essence of the approach is that the boundary conditions are applied during formulating the weighting coefficient matrices for inner grid points. The technique resulted in an obvious improvement in the DQ solution of beams and plates with free and SS boundary conditions [12, 21, 22]. However, it is regret that the technique is not applicable for problems with the C-C boundary condition as well as cross derivative and corner boundary conditions such as completely free plates. Thus, a combination of the DQWB and DQδ approaches was used to handle the problems with both the C-C boundary condition and the other boundary conditions in [22]. The modification of the weighting coefficient matrices also causes some loss of use flexibility and increases rather high additional computational effort, which require some matrix product operations of $O(N^4)$ scalar multiplications. In this study, the DQWB approach is used to handle the problems with simple supports.

Chen and Yu [23] proposed a different DQCY approach to cure the deficiencies of the conventional DQδ approach. The fact was noted that that the rank of the DQ weighting coefficient matrix for the i-th order derivative is N-i, where N is the number of grid points. Naturally, the rank of the DQ coefficient matrix for the $4^{th}$ order derivative is N-4. Therefore, the DQ analog equations of the $4^{th}$ order governing equations at boundary points itself and the points immediate adjacent boundary need be replaced by the analog equations of the boundary equations. DQCY approach imposes all boundary conditions exactly at boundary points. Therefore, the solution accuracy of the DQCY approach is improved evidently in comparison to the DQδ approach. The approach is conceptually simple and effective for the C-C boundary condition as well as any other boundary condition. Since the DQCY eliminates the δ effect in the DQδ, the stability of the solution is improved and larger number of grid points can be used. However, it has found that the DQCY approach is less efficiency than the DQWB approach whenever the DQWB approach is applicable. Therefore, in this paper the DQCY approach is applied to handle the problems with the C-C boundary condition.

An improvement is also made here for the DQCY approach. In this application, the first and last two rows of the original weighting coefficient matrices are removed, and then the four boundary equations are applied to modify these coefficient matrices into (N-4)×(N-4) matrix of full rank. The modified weighting coefficient matrices can be used to formulate the differential equations of interest directly. Since the boundary conditions are incorporated into the DQ weighting coefficient matrices in advance, extra analog equations at four corner points of plate need no longer be considered in the resulting formulation. The size of the resultant algebraic equations is also reduced. For example, consider the C-C boundary:

$w = 0$,  at x=0, 1; (2-7a)

$\frac{\partial w}{\partial x} = 0$,  at x=0, 1. (2-7b)

The corresponding DQ approximate equations are given by

$$w_1 = 0, \quad w_N = 0; \tag{2-8a}$$

$$\sum_{j=1}^{N} A_{1j} w_j = 0, \quad \sum_{j=1}^{N} A_{Nj} w_j = 0, \tag{2-8b}$$

where $w_j$ is the displacement at the jth grid point. It is noted that all boundary conditions are exactly satisfied at boundary points. By using equation (2-8a, b), the desired displacements at the 2nd and (N-1)th grid points can be expressed in terms of the unknown displacement values at interior points, namely,

$$w_2 = -\frac{1}{A_{12}} \sum_{j=3}^{N-1} A_{1j} w_j \tag{2-9a}$$

$$w_{N-1} = -\frac{1}{A_{N,N-1}} \sum_{j=2}^{N-2} A_{Nj} w_j \tag{2-9b}$$

Substituting equations (2-8a) and (2-9a, b) into the DQ approximate formulas for the 1st, 2nd, 3rd and 4th derivatives, respectively, we have

$$\frac{d\bar{w}}{dx} = \bar{A}\bar{w}, \quad \frac{d^2\bar{w}}{dx^2} = \bar{B}\bar{w}, \quad \frac{d^3\bar{w}}{dx^3} = \bar{C}\bar{w}, \quad \frac{d^4\bar{w}}{dx^4} = \bar{D}\bar{w} \tag{2-10}$$

where $\bar{w} = \{w_3, w_4, \ldots, w_{N-2}\}$. $\bar{A}$, $\bar{B}$, $\bar{C}$ and $\bar{D}$ are (N-4)×(N-4) modified coefficient matrices different from those of (N-2)×(N-2) dimension in the DQWB approach. For the other boundary conditions, the modified coefficient matrices can be obtained in the similar way.

## 2.2 Matrix approximate formulas in the DQ method

One of the present authors [20] proposed the following DQ formulation in matrix form for the partial derivatives of the function $\psi(x,y)$ in two-dimensional domain:

$$\frac{\partial \hat{\psi}}{\partial x} = A_x \hat{\psi}, \quad \frac{\partial \hat{\psi}}{\partial y} = \hat{\psi} A_y^T, \quad \frac{\partial^2 \hat{\psi}}{\partial x \partial y} = A_x \hat{\psi} A_y^T, \quad \frac{\partial^2 \hat{\psi}}{\partial x^2} = B_x \hat{\psi},$$

$$\frac{\partial^2 \hat{\psi}}{\partial y^2} = \hat{\psi} B_y^T, \quad \frac{\partial^2 \hat{\psi}}{\partial x \partial y} = B_x \hat{\psi} B_y^T, \quad \frac{\partial^2 \hat{\psi}}{\partial x^2} = D_x \hat{\psi}, \quad \frac{\partial^2 \hat{\psi}}{\partial y^2} = \hat{\psi} D_y^T, \tag{2-11}$$

where the unknown $\hat{\psi}$ is a rectangular unknown matrix rather than a vector as in [4, 6, 8, 11-14, 21, 22]. $\bar{A}$, $\bar{B}$ and $\bar{D}$ with subscripts x and y stand for the DQ weighting coefficient matrices for the 1st, 2nd and 4th order partial derivatives along x and y directions, respectively. The superscript T means the transpose of the matrices. The above analog equations are compact matrix version of the traditional polynomial approximate formulas given by Civan and Sliepcevich [4]. In fact, both are equivalent.

### 3. Hadamard and SJT product

It is not an easy task to handle the problems involving nonlinearity. Chen and Zhong [15] first found that the Hadamard product of matrices was a simple and effective technique to formulate nonlinear differential equations in the DQ method. The SJT product was also introduced there to efficiently compute the Jacobian derivative matrix in the Newton-Raphson method for the solution of nonlinear formulation in the Hadamard product form.

**Definition 3.1** Let matrices $A=[a_{ij}]$ and $B=[b_{ij}] \in C^{N \times M}$, the Hadamard product of matrices is defined as $A \circ B = [a_{ij} b_{ij}] \in C^{N \times M}$, where $C^{N \times M}$ denotes the set of N×M real matrices.

**Definition 3.2** If matrix $A=[a_{ij}] \in C^{N \times M}$, then $A^{\circ q}=[a_{ij}^q] \in C^{N \times M}$ is defined as the Hadamard power of matrix A, where q is a real number. Especially, if $a_{ij} \neq 0$, $A^{\circ(-1)}=[1/a_{ij}] \in C^{N \times M}$ is defined as the Hadamard inverse of matrix A. $A^{\circ 0}=11$ is defined as the Hadamard unit matrix in which all elements are equal to unity.

**Definition 3.3** If matrix $A=[a_{ij}] \in C^{N \times M}$, then the Hadamard matrix function $f^{\circ}(A)$ is defined as $f^{\circ}(A) = [f(a_{ij})] \in C^{N \times M}$.

Considering quadratic nonlinear differential operator $W_{,x}W_{,y}$, the DQ formulation can be expressed in Hadamard product form as

$$W_{,x}W_{,y} = (A_x \vec{W}) \circ (A_y \vec{W}),  \qquad (3\text{-}1)$$

where $W$ is the unknown vector. $W$ with the comma subscript ,x and ,y denotes vector of partial derivative of function W along x- and y- directions. $\overline{A_x}$ and $\overline{A_y}$ are the DQ weighting coefficient matrices, modified by the boundary conditions, for the 1st order derivative along the x- and y- directions, respectively. The above equation exposes an essential idea in the latter analysis. For linear and nonlinear operators with varying parameters c(x,y), the DQ analog can be obtained in the same way. For example,

$$c(x,y)U_{,x} = \{c(x_j,y_j)\} \circ (\overline{A_x}\overline{U}) \tag{3-2a}$$

and

$$c(x,y)U_{,x}U_{,y} = \{c(x_j, y_j)\} \circ (\overline{A_x}U) \circ (\overline{A_y}U). \tag{3-2b}$$

**Theorem 3.1**: If A, B and C $\in C^{N \times M}$, then

1> A°B=B°A  (3-3a)

2> k(A°B)=(kA)°B, where k is a scalar.  (3-3b)

3> (A+B)°C=A°C+B°C  (3-3c)

The Newton-Raphson method may be one of the most important techniques to compute nonlinear algebraic equations. One of the major time-consuming calculations in the method is to evaluate the Jacobian derivative matrix. In order to simplify the computation of the Jacobian matrix, the postmultiplying SJT product of matrix and vector was introduced by Chen and Zhong [15].

**Definition 3.4** If matrix $A=[a_{ij}] \in C^{N \times M}$, vector $\vec{V} = \{v_j\} \in C^{N \times 1}$, then $A \lozenge \vec{V} =[a_{ij}v_j] \in C^{N \times M}$ is defined as the postmultiplying SJT product of matrix A and vector $\vec{V}$, where ◊ represents the SJT product.

In what follows we give the operation rules applying the SJT product to compute the Jacobian matrices, where $\frac{\partial}{\partial \bar{U}}$ denotes the Jacobian derivative matrix operator with respect to vector $\bar{U}$.

1. For $c(x,y)U_{,x} = \{c(x_j,y_j)\} \circ (\overline{A_x}\bar{U})$, we have

$$\frac{\partial}{\partial \bar{U}}\{\{c(x_j,y_j)\} \circ (\overline{A_x}\bar{U})\} = \overline{A_x} \Diamond \{c(x_j,y_j)\}. \tag{3-4a}$$

2. For $(U_{,x})^q = (\overline{A_x}\bar{U})^{\circ q}$, where q is a real number, we have

$$\frac{\partial}{\partial \bar{U}}\{(\overline{A_x}\bar{U})^{\circ q}\} = q\overline{A_x} \Diamond (\overline{A_x}\bar{U})^{\circ(q-1)}. \tag{3-4b}$$

3. For $U_{,x}U_{,y} = (\overline{A_x}\bar{U}) \circ (\overline{A_y}\bar{U})$, we have

$$\frac{\partial}{\partial \bar{U}}\{(\overline{A_x}\bar{U}) \circ (\overline{A_y}\bar{U})\} = \overline{A_x} \Diamond (\overline{A_y}\bar{U}) + \overline{A_y} \Diamond (\overline{A_x}\bar{U}) \tag{3-4c}$$

4. For $\sin U_{,x} = \sin {}^{\circ}(\overline{A_x}\bar{U})$, we have

$$\frac{\partial}{\partial \bar{U}}\{\sin(\overline{A_x}\bar{U})\} = \overline{A_x} \Diamond \cos{}^{\circ}(\overline{A_x}\bar{U}) \tag{3-4d}$$

5. For $\exp(U_{,x}) = \exp{}^{\circ}(\overline{A_x}\bar{U})$, we have

$$\frac{\partial}{\partial \bar{U}}\{\exp{}^{\circ}(\overline{A_x}\bar{U})\} = \overline{A_x} \Diamond \exp{}^{\circ}(\overline{A_x}\bar{U}) \tag{3-4e}$$

The above computing formulas give the analytic solution of the Jacobian derivative matrix for the analog equations considered. The computational effort for one SJT product is only $n^2$ scalar multiplications, which may be the smallest computational cost in all possible approaches. The premultiplying SJT product was also introduced to compute the Jacobian matrix of the nonlinear formulation such as $\frac{\partial \bar{W}^m}{\partial x} = \overline{A_x}\bar{W}^m$ ($m \neq 1$) [15, 16]. Such cases are seldom encountered in structure analysis. Therefore, it is not presented here for the sake of brevity.

The finite difference method is a simple technique to obtain the approximate solution of the Jacobian matrix in practical engineering and requires $O(n^2)$ scalar multiplications.

Both the SJT product approach and the finite difference method are essentially comparable in computing effort. However, the approximate Jacobian matrix yielded by the finite difference method affects the accuracy and convergence rate of the Newton-Raphson method. In contrast, the SJT product produces the analytic solution of the Jacobian matrix.

## 4. APPLICATIONS

The von Karman equations governing a thin, homogeneous, orthotropic rectangular plate subject to a uniformly distributed transverse load are given by [9]

$$E_1 u_{,xx} + \mu G_{12} u_{,yy} + C v_{,xy} = -w_{,x}\left(E_1 w_{,xx} + \mu G_{12} w_{,yy}\right) - C w_{,y} w_{,xy} \qquad (4\text{-}1a)$$

$$E_2 v_{,yy} + \mu G_{12} v_{,xx} + C u_{,xy} = -w_{,y}\left(E_2 w_{,yy} + \mu G_{12} w_{,xx}\right) - C w_{,x} w_{,xy} \qquad (4\text{-}1b)$$

$$\begin{aligned} D_1 w_{,xxxx} + 2 D_3 w_{,xxyy} + D_2 w_{,yyyy} &= q + \frac{h}{\mu}\Bigg[\left(u_{,x} + \frac{1}{2}w_{,x}^2\right)\left(E_1 w_{,xx} + \upsilon_{12} E_2 w_{,yy}\right) \\ &+ \left(v_{,y} + \frac{1}{2}w_{,y}^2\right)\left(E_2 w_{,yy} + \upsilon_{21} E_1 w_{,xx}\right) + 2\mu G_{12} w_{,xy}\left(u_{,y} + v_{,x} + w_{,x} w_{,y}\right)\Bigg] \end{aligned} \qquad (4\text{-}1c)$$

in terms of three displacement components, where $\nu_{12}$ and $\nu_{21}$ are Poisson's ratio, $E_1$ and $E_2$ are the Young's moduli. C is the shear modulus, $D_1$, $D_2$ and $D_4$ are the principal bending and twisting rigidities, $\mu=1-\nu_{12}\nu_{21}$, u, v and w are the desired inplane and transverse displacements. a, b and h are width, length and thickness of plate, respectively.

Applying the Hadamard product and the new DQ matrix approximate formulas (2-11), the analog equations for this case are

$$\begin{aligned} E_1 \overline{B_x}\hat{u} + \mu G_{12}\hat{u}\overline{B_y^T} + C\overline{A_x}\hat{v}\overline{A_y^T} &= -\left(\overline{A_x}\hat{w}\right)\circ\left(E_1 \overline{B_x}\hat{w} + \mu G_{12}\hat{w}\overline{B_y^T}\right) \\ &\quad - C\left(\hat{w}\overline{A_y^T}\right)\circ\left(\overline{A_x}\hat{w}\overline{A_y^T}\right) \end{aligned} \qquad (4\text{-}2a)$$

$$\begin{aligned} E_2 \hat{v}\overline{B_y^T} + \mu G_{12}\overline{B_x}\hat{v} + C\overline{A_x}\hat{u}\overline{A_y^T} &= -\left(\hat{w}\overline{A_y^T}\right)\circ\left(E_2 \hat{w}\overline{B_y^T} + \mu G_{12}\overline{B_x}\hat{w}\right) \\ &\quad - C\left(\overline{A_x}\hat{w}\right)\circ\left(\overline{A_x}\hat{w}\overline{A_y^T}\right) \end{aligned} \qquad (4\text{-}2b)$$

$$D_1\overline{D}_x\hat{w}+2D_3\overline{B}_x\hat{w}\overline{B}_y^T+D_2\hat{w}\overline{D}_y^T=q+\frac{h}{\mu}\left[\left[\overline{A}_x\hat{u}+\frac{1}{2}(\overline{A}_x\hat{w})^{\circ 2}\right]\right.$$

$$\circ\left[E_1\overline{B}_x\hat{w}+\upsilon_{12}E_2\hat{w}\overline{B}_y^T\right]+\left[\hat{v}\overline{A}_y^T+\frac{1}{2}(\hat{w}\overline{A}_y^T)^{\circ 2}\right]\circ\left[E_2\hat{w}\overline{B}_y^T+\upsilon_{21}E_1\overline{B}_x\hat{w}\right] \quad (4\text{-}2c)$$

$$\left.+2\mu G_{12}(\overline{A}_x\hat{w}\overline{A}_y^T)\circ\left[\hat{u}\overline{A}_y^T+\overline{A}_x\hat{v}+(\overline{A}_x\hat{w})\circ(\hat{w}\overline{A}_y^T)\right]\right]$$

where $\overline{A}$, $\overline{B}$ and $\overline{D}$ with subscript x and y denote the modified weighting coefficient matrix along x and y directions, respectively. The orders of these matrices are N-2 for the DQWB approach and N-4 for the DQCY approach, where N is the number of grid points. Note that the boundary conditions have been applied in the DQWB and DQCY approach and thus are no longer considered. It is also noted that $\overline{A}_x$ and $\overline{A}_y$ here are different form those in section 3. The latter are the stacked coefficient matrix, the corresponding $W$ and $U$ are the desired vectors, while the former are in one-dimensional sense and $\hat{u}$, $\hat{v}$ and $\hat{w}$ in the above formulations are rectangular matrices. It is also pointed out that $\overline{A}_x$, $\overline{B}_x$, $\overline{A}_y$ and $\overline{B}_y$ for different u, v and w are the same, respectively, in the cases of clamped or simply supported edges.

The variables are nondimensionalized as $X\equiv x/a, Y\equiv y/b, U\equiv \hat{u}/a, V\equiv \hat{v}/b$ and $W\equiv \hat{w}/h$. Furthermore, by using the Kronecker product of matrices, we have

$$H_1\vec{U}+H_2\vec{V}=-(H_7\vec{W})\circ(H_1\vec{W})-(H_8\vec{W})\circ(H_2\vec{W}) \quad (4\text{-}3a)$$

$$H_2\vec{U}+H_3\vec{V}=-(H_8\vec{W})\circ(H_3\vec{W})-(H_7\vec{W})\circ(H_2\vec{W}) \quad (4\text{-}3b)$$

$$H_4\vec{W}=\frac{qa^4}{D_1h}+\frac{a^4}{\mu D_1h}\left[\frac{a^2}{h^2}\left[H_7\vec{U}+\frac{1}{2}(H_7\vec{W})^{\circ 2}\right]\circ[H_5\vec{W}]+\frac{b^2}{h^2}\left[H_8\vec{V}+\frac{1}{2}(H_8\vec{W})^{\circ 2}\right]\right.$$

$$\left.\circ[H_6\vec{W}]+\frac{2\mu G_{12}}{C}(H_2\vec{W})\circ[H_8\vec{U}+H_7\vec{V}+(H_7\vec{W})\circ(H_8\vec{W})]\right] \quad (4\text{-}3c)$$

where $H_1$, $H_2$, $H_3$, $H_4$, $H_5$, $H_6$, $H_7$ and $H_8$ see appendix. $\vec{W}$, $\vec{U}$ and $\vec{V}$ are vectors generated by stacking the rows of the corresponding rectangular matrix W, U, and V into one long vector. By using the new matrix approximate formulas and Hadamard product, the present nonlinear formulations are very easily accomplished. The matrix form here is also simpler and more explicit than the conventional algebraic polynomial form given by Bert et al. [9].

Equations (4-3a) and (4-3b) can be also restated as

$$H_1 \bar{U} + H_2 \bar{V} = -L_1(\bar{W}) \tag{4-4a}$$

$$H_2 \bar{U} + H_3 \bar{V} = -L_2(\bar{W}), \tag{4-4b}$$

where

$$L_1(\bar{W}) = (H_7 \bar{W}) \circ (H_1 \bar{W}) + (H_8 \bar{W}) \circ (H_2 \bar{W}) \tag{4-5a}$$

$$L_2(\bar{W}) = (H_8 \bar{W}) \circ (H_3 \bar{W}) + (H_7 \bar{W}) \circ (H_2 \bar{W}). \tag{4-5b}$$

The unknown vector $\bar{U}$ and $\bar{V}$ can be expressed in terms of $\bar{W}$ by

$$\bar{U} = H_9^{-1} H_3^{-1} L_2(\bar{W}) - H_9^{-1} H_2^{-1} L_1(\bar{W}) \tag{4-6a}$$

and

$$\bar{V} = H_{10}^{-1} H_2^{-1} L_2(\bar{W}) - H_{10}^{-1} H_1^{-1} L_1(\bar{W}), \tag{4-6b}$$

where $H_9 = H_2^{-1} H_1 - H_3^{-1} H_2$ and $H_{10} = H_1^{-1} H_2 - H_2^{-1} H_3$. By applying the SJT product for the evaluation of the Jacobian matrix, we have

$$\frac{\partial \bar{U}}{\partial \bar{W}} = H_9^{-1} H_3^{-1} \frac{\partial L_2(\bar{W})}{\partial \bar{W}} - H_9^{-1} H_2^{-1} \frac{\partial L_1(\bar{W})}{\partial \bar{W}} \tag{4-7a}$$

and

$$\frac{\partial \bar{V}}{\partial \bar{W}} = H_{10}^{-1} H_2^{-1} \frac{\partial L_2(\bar{W})}{\partial \bar{W}} - H_{10}^{-1} H_1^{-1} \frac{\partial L_1(\bar{W})}{\partial \bar{W}}, \tag{4-7b}$$

where

$$\frac{\partial L_1(\bar{W})}{\partial \bar{W}} = H_7 \Diamond (H_1 \bar{W}) + H_1 \Diamond (H_7 \bar{W}) + H_8 \Diamond (H_2 \bar{W}) + H_2 \Diamond (H_8 \bar{W}) \tag{4-8a}$$

and

$$\frac{\partial L_2(\bar{W})}{\partial \bar{W}} = H_8 \Diamond (H_3 \bar{W}) + H_3 \Diamond (H_8 \bar{W}) + H_7 \Diamond (H_2 \bar{W}) + H_2 \Diamond (H_7 \bar{W}). \tag{4-8b}$$

$\dfrac{\partial \vec{U}}{\partial \vec{W}}$ and $\dfrac{\partial \vec{V}}{\partial \vec{W}}$ are relative Jacobian derivative matrices of dependent variable vector $\vec{U}$ and $\vec{V}$ with respect to $\vec{W}$. Applying formulas (4-6a, b) and (4-7a, b), the coupled formulations (4-3a, b, c) are decoupled. The size of the resulting nonlinear simultaneous algebraic equations are reduced from 3(N-2)×3(N-2) to (N-2)×(N-2) for the DQWB approach or from 3(N-4)×3(N-4) to (N-4)×(N-4) for the DQCY approach. It is known that each iteration step in the Newton-Raphson method has to solve a set of linear simultaneous algebraic equations, which requires an order of $n^3$ scalar multiplications. n is the size of the solved equations. For example, the Gauss elimination method requires $n^3/3$ scalar multiplications, while the Gauss-Jordan method requires $n^3/2$ scalar multiplications. Therefore, the computational effort and storage requirements here are only about one twenty-seventh and one-ninth, respectively, as much as in reference [9]. $\vec{W}$ is a basic variable vector in the present computations. Equation (4-3c) is chosen as the basic iteration equation, namely,

$$\varphi\{\vec{W}\} = H_4 \vec{W} - \dfrac{a^4}{\mu D_1 h} \left[ \dfrac{a^2}{h^2} \left[ H_7 \vec{U} + \dfrac{1}{2}(H_7 \vec{W})^{\circ 2} \right] \circ [H_5 \vec{W}] + \dfrac{b^2}{h^2} \left[ H_8 \vec{V} + \dfrac{1}{2}(H_8 \vec{W})^{\circ 2} \right] \right.$$
$$\left. \circ [H_6 \vec{W}] + \dfrac{2\mu G_{12}}{C}(H_2 \vec{W}) \circ [H_8 \vec{U} + H_7 \vec{V} + (H_7 \vec{W}) \circ (H_8 \vec{W})] \right] - \dfrac{qa^4}{D_1 h} = 0 \quad (4\text{-}9)$$

The Jacobian derivative matrix for the above iteration equation is given by

$$\dfrac{\partial \varphi\{\vec{W}\}}{\partial \vec{W}} = H_4 - \dfrac{a^4}{\mu D_1 h} \left[ \dfrac{a^2}{h^2} \left[ H_7 \dfrac{\partial \vec{U}}{\partial \vec{W}} + H_7 \Diamond (H_7 \vec{W}) \right] \Diamond [H_5 \vec{W}] \right.$$
$$+ \dfrac{a^2}{h^2} H_5 \Diamond \left[ H_7 \vec{U} + \dfrac{1}{2}(H_7 \vec{W})^{\circ 2} \right] + \dfrac{b^2}{h^2} \left[ H_8 \dfrac{\partial \vec{V}}{\partial \vec{W}} + H_8 \Diamond (H_8 \vec{W}) \right] \Diamond [H_6 \vec{W}]$$
$$+ \dfrac{b^2}{h^2} H_6 \Diamond \left[ H_8 \vec{U} + \dfrac{1}{2}(H_8 \vec{W})^{\circ 2} \right] + \dfrac{2\mu G_{12}}{C} H_2 \Diamond [H_8 \vec{U} + H_7 \vec{V} + (H_7 \vec{W}) \circ (H_8 \vec{W})] \quad (4\text{-}10)$$
$$\left. + \dfrac{2\mu G_{12}}{C} \left[ H_8 \dfrac{\partial \vec{U}}{\partial \vec{W}} + H_7 \dfrac{\partial \vec{V}}{\partial \vec{W}} + H_7 \Diamond (H_8 \vec{W}) + H_8 \Diamond (H_7 \vec{W}) \right] \Diamond (H_2 \vec{W}) \right]$$

It is noted that the SJT product approach here yields the analytical solution of the Jacobian matrix quite simply and efficiently. The Newton-Raphson iteration equation for this case is

$$\bar{W}^{(k+1)} = \bar{W}^{(k)} - \left(\frac{\partial \varphi(\bar{W}^{(k)})}{\partial \bar{W}}\right)^{-1} \varphi(\bar{W}^{(k)}) \qquad (4\text{-}11)$$

The DQ solutions for the corresponding linear isotropic and orthotropic plates are chosen as the initial guess of the iteration procedures. It is noted that the Newton-Raphson method has rather big convergence domain in the present computations. Even if the resulting nonlinear results are even eight times larger than the initial linear solutions, the Newton-Raphson method still converges. Moreover, the solution converges very rapidly, and the iterative times varies from 1 to 10 for various loadings when the convergence criterion is the maximum residuals of equation (4-9) no more than $10^{-5}$. In contrast, the IMSL subroutine NEONE used in the reference [9] computed the Jacobian matrix approximately by a finite difference technique. Therefore, as was shown in Fig. 8 in reference [9], the accuracy and converging rate of the DQ solutions are affected, especially for simply supported plates.

References [8, 12, 17] pointed out that the accuracy and stability of the DQ method can be improved significantly if the Chebyshev grid spacing is used. In the following the DQWB and DQCY solutions are obtained by using Chebyshev grid spacing 7×7 for a simply supported plate and 9×9 for a clamped plate, respectively, unless where specified. To avoid the effects of round-off errors on the accuracy of the solution, double-precision arithmetic is used in all the results presented in this paper. Reference [9] has pointed out the high efficiency and ease of use in the DQ method in comparison to other numerical techniques such as the finite element, finite difference, perturbation, Galerkin and Rayleigh-Ritz, etc., while this paper places its emphasis in the simplification of the use and further improvement of the accuracy and efficiency in the DQ method. Therefore, the comparisons with the other numerical techniques are not discussed here.

The same simply-supported and clamped isotropic square plates subject to a uniformly distributed loading as in example 1 of reference [9] are recalculated by the present DQ method. The solutions by both the DQCY and conventional DQδ approaches are nearly agreement and very accurate for the clamped plate, and, thus, the DQCY solutions are not presented here for the sake of brevity. Nevertheless, it should be emphasized that the DQδ approach can not use larger number of grid points as in the DQCY approach due to instability caused by the δ effect. The results for simply supported plate are shown in Fig. 1 and compared with the exact [24, 25] and the conventional DQ [9] solutions. The present DQWB results show remarkable agreement with those of Levy [24] and Yang [25]. It is also noted that the DQδ approach using 7×7 grid points by Bert et al. [9] gave obviously better results in the clamped cases than in the simply supported cases as shown in Figs 1 and 2 of reference [9]. This is because the DQδ approach is not very suitable for the cases of supported edges. As is expected, the DQWB approach gives more accurate results than the DQδ approach for simply supported plate. Therefore, the former is a competitive alternative to the latter for the nonlinear cases of simple supports.

In addition, we compute an isotropic simply supported square plate under a uniformly distributed transverse load. The parameters of this case are a=16", h=0.1", E=30E+6 and ν=0.316. Fig. 2 depicts the results obtained by the DQWB approach using 5×5 Chebyshev grids and 7×7 equally spaced grids. All solutions agree very well with those given by Levy [26]. As is expected, the DQWB method using the 5×5 Chebyshev grids yields more accurate results than using 7×7 equally spaced grids.

The center deflections of the clamped square plate (a=100, h=1.0, E=2.1E+6, ν=0.316, q=3.0) and the simply-supported square plate (a=100, h=1.0, E=2.1E+6, ν=0.25, q=1.0) subject to a uniformly distributed pressure are computed, respectively, by the DQWB and DQCY methods. The DQ results as well as the analytical and FEM solutions are listed in Table 1. The present DQ solutions show excellent agreement with the analytical [27] and FEM solutions [27, 28]. However, the computational effort in the present DQ method is much less than in the analytical method and FEM. The DQ method is demonstrated again to be highly computationally efficient for nonlinear structural analysis.

The numerical examples on the orthotropic rectangular plate provided by Bert et al. [9] are recalculated by the present DQ methods. The specific parameters are $E_1$=18.7E+6psi; $E_2$=1.3E+6psi, $G_{12}$=0.6E+6psi; $\nu_{12}$=0.3; h=0.0624inch; a=9.4inch; b=7.75inch. The center deflections in the clamped and simply supported cases are displayed in Figs. 3 and 4, respectively. For the cases of clamped edges, the results by the DQCY approach are very close to those by Bert and Cho. It is noted that the DQCY approach using 15×15 grid spacing is still stable and give accurate results, but computational effort also increases exponentially, while the DQδ approach can not use so many grid points. The computational stability problem may be essential in some cases when many grid points are required. Therefore, the DQCY approach may have better promise in practical engineering than the DQδ approach. In this case, the DQCY approach produces more accurate results by using more grid points (e.g., 9×9 grid spacing).

For the case of simple supports, it is noted that the results given by Bert, Striz and Jang [9] using DQδ approach with 7×7 equally spaced grids are obviously larger than those by Bert and Cho. In contrast, the present DQWB approach appears to give results that are much closer to those by Bert and Cho as shown in Fig. 4. Bert and Cho's solution values here are taken from graphs 9 and 10 in respective reference [9] with appropriate scaling factors. Another example on orthotropic plates discussed in reference [9] are also recalculated by the present DQ method and the same conclusions are obtained.

Based on the centrosymmetric structures of the DQ weighting coefficient matrices, the reduction technique in the DQ method was proposed in [16, 17, 20] and extended to the nonlinear problems in [16]. For geometrically nonlinear bending of the isotropic and orthotropic rectangular plates with symmetric boundary conditions, the computational effort and storage requirements can be further reduced by 75% and 50% using such reduction technique, respectively.

## 5. CONCLUDING REMARKS

The DQ approach in using some new techniques is applied to analyze geometrically nonlinear bending of isotropic and orthotropic plates with simply supported and clamped edges. The new matrix approximate formulas offer a compact and convenient DQ procedure to approximate partial derivatives. The DQWB approach is proved to be a successful technique for geometrically nonlinear plate with SS-SS boundary conditions. It is conceivable that the DQWB approach is also highly efficient for the other boundary conditions whenever applicable as in the linear problems. The DQCY approach is improved and shown to be a stable and accurate technique for handling the cases with the C-C boundary conditions. Apparently, the results obtained by these two approaches are more accurate than those by the traditional DQ$\delta$ approach. Although only simply-supported and clamped boundary conditions are involved in the present study, it is straightforward that the same procedures can be easily employed for handling problems with the other boundary conditions.

The publications in which the DQ method was applied to deal with nonlinear problems are still few due to much more complex programming, storage requirements, formulation and computing effort in comparison to linear problems. The Hadamard and SJT product approach may provide a simple and efficient technique to greatly reduce these impediments. The detailed solution procedures for the geometrically nonlinear plate cases are provided here to show the simplicity and efficiency of the Hadamard and SJT product approach. It is worth stressing that the Hadamard and SJT products as well as the relative decoupling technique are applicable for the finite difference, pseudo-spectral and collocation methods. Also, although only the application of Hadamard and SJT product within one single interval is presented in this paper, the extension of this procedure to solutions of problems with complex geometries with the coordinate mappings and multidomain techniques should be straightforward. The application of the present DQ solution procedure to the nonlinear analysis of more complex plates with varying thickness, Poisson's ratio and Young's modulus is currently the subject of further investigation.

## REFERENCES

1. R. E. Bellman and J. Casti, Differential quadrature and long-term integration. J. Math. Anal. Appl. 34, 235-238 (1971).

2. R. E. Bellman, B. G. Kashef and J. Casti, Differential quadrature: A technique for the rapid solution of nonlinear partial differential equations. J. Comput. Phys.10, 40-52 (1972).

3. J. O. Mingle, The method of differential quadrature for transient nonlinear diffusion. J. Math. Anal. Appl. 60, 559-569 (1977).

4. F. Civan and C. M. Sliepcevich, Differential quadrature for multidimensional problems. J. Math. Anal. Appl. 101, 423-443 (1984).

5. G. Naadimuthu, R. E. Bellman, K. M. Wang and E. S. Lee, Differential quadrature and partial differential equations: some numerical results. J. Math. Anal. Appl. 98, 220-235 (1984).

6. C. W. Bert, S. K. Jang and A. G. Striz, Two new methods for analyzing free vibration of structure components. AIAA J. 26, 612-618 (1988).

7. A. G. Striz, S. K. Jang and C. W. Bert, Nonlinear bending analysis of thin circular plates by differential quadrature. Thin-Walled Struct. 6, 51-62 (1988).

8. J. R. Quan and C. T. Chang, New insights in solving distributed system equations by the quadrature methods - 2: Numerical Experiments. Comput. Chem. Engrg. 13, 1017-1024 (1989).

9. C. W. Bert, A. G. Striz and S. K. Jang, Nonlinear bending analysis of orthotropic rectangular plates by the method of differential quadrature. Comput. Mech. 5, 217-226 (1989).

10. Y. Feng and C. W. Bert, Application of the quadrature method to flexural vibration analysis of a geometrically nonlinear beam, Nonlinear Dynamics. 3, 13-18 (1992).

11. C. Shu and B. E. Richards, Parallel simulation of incompressible viscous flows by generalized differential quadrature. Computing Systems in Engrg. 3, 271-281 (1992).

12. C. W. Bert, X. Wang and A. G. Striz, Differential quadrature for static and free vibrational analysis of anisotropic plates. Int. J. Solids Struct. 30, 1737-1744 (1993).

13. P. A. A. Laura and R. E. Rossi, "The method of differential quadrature and its application to the approximate solution of ocean engineering problems," Ocean Eng. 21, 57-66(1994).

14. M. Malik and F. Civan, A comparative study of differential quadrature and cubature methods vis-à-vis some conventional techniques in context of convection-diffusion-reaction problems. Chem. Engrg. Sci. 50, 531-547 (1994).

15. Wen Chen and Tingxiu Zhong, The study on nonlinear computations of the DQ and DC methods. Numer. Methods for Partial Differential Equations 13, 57-75 (1997).

16. Wen Chen, Differential Quadrature method and its applications in Engineering, Ph.D. dissertation, Shanghai Jiao Tong University, China, 1996.

17. J. R. Quan and C. T. Chang, New insights in solving distributed system equations by the quadrature methods - 1. Comput. Chem. Engrg 13, 779-788 (1989).

18. G. Mansell, W. Merryfield, B. Shizgal and U. Weinerl, A comparison of differential quadrature methods for the solution of partial differential equations, Comput. Methods Appl. Mech. Engrg. 104, 295-316 (1993).

19. M. K. Burka, Solution of stiff ordinary differential equations by decomposition and orthogonal collocation. AIChE J. 28, 11-20 (1982).

20. Wen Chen, X, Wang and Y. Yong, Reducing the computational effort of the differential quadrature method, Numerical Methods for Partial Differential Equations 12, 565-577 (1996).

21. X. Wang and C. W. Bert, A new approach in applying differential quadrature to static and free vibrational analyses of beams and plates. J. Sound & Vibr. 162, 566-572 (1993).

22. X. Wang, C. W. Bert and A. G. Striz, Differential quadrature analysis of deflection, buckling and free vibration of beams and rectangular plates. Comput. Struct. 48, 473-479 (1993).

23. Wen Chen and Y. Yu, Differential quadrature method for high order boundary value problems. In: Proc. 1st Pan-Pacific Conf. on Comput. Engrg. (Kwak, B. M., Tanaka, M. eds.). Elsevier Sci. Publ. B. V. Netherlands, pp. 163-168 (1993).

24. T. Y. Yang, Finite displacement plate flexure by the use of matrix incremental approach, Int. J. Numer. Meths. Engrg. 4, 415-432 (1972).

25. S. Levy, Square plate with clamped edges under normal pressure producing large deflections. NACA Report No. 740 (1942).

26. S. Levy, Bending of rectangular plates with large deflections. NACA Report No. 737. 139-157 (1942).

27. G. Zhu and H. Wang, Quasi-conforming penalty FEM for large deflection of composite laminated plate (in Chinese). ACTA MATERIAE COMPOSITE SINICA, 6, 39-47 (1989).

28. G. P. Bazeley, Y. K. Cheung, B. M. Irens and O. C. Zienkiewicz, Triangular elements in bending-conforming and non-conforming solutions, Proc. Conf. Matrix Methods in Structural Mechanics, J. S. Przemieniecki et al. eds. Chio: Air Force Ins. Tech. Wright-Patterson A F Base. 547-576 (1965).

29. P. Lancaster and M. Timenetsky, The Theory of Matrices with Applications, 2nd edition, Academic Press. Orlando (1985).

**APPENDIX**

The desired unknowns in rectangular matrix form as in equations (4-2a, b, c) can be converted into the conventional vector form by using the Kronecker product [29].

**Lemma 1.** If $A \in C^{p \times m}$, $B \in C^{n \times q}$ and the unknown $X \in C^{m \times n}$, then

$$vec(AXB) = (A \otimes B^T)vec(X) \tag{A1}$$

where vec( ) is the vector-function of a rectangular matrix formed by stacking the rows of matrix into one long vector, $\otimes$ denotes the Kronecker product of matrices. Note that vec( ) here is to stack rows rather than columns as in reference [29] and, thus, the corresponding operation rules are also slightly different, but both are in fact equivalent. In this paper, we define $vec(AXB) = A X \bar{B}$ and $vec(X) = \bar{X}$ to simplify presentation.

Corollary:

$$\begin{aligned} &1.\, A\bar{X} = (A \otimes I_n)\bar{X} \\ &2.\, X\bar{B} = (I_m \otimes B^T)\bar{X} \\ &3.\, A\bar{X} + X\bar{B} = (A \otimes I_n + I_m \otimes B^T)\bar{X} \end{aligned} \tag{A2}$$

where $I_n$ and $I_m$ are the unit matrix. According to the above formulas, Eqs. (4-2a, b, c) can be converted into Eqs. (4-3a, b, c), and the resulting coefficient matrices in Eqs. (4-3a, b, c) are given by

$$H_1 = E_1\left(\bar{B}_x \otimes I_y\right) + \mu G_{12}\left(\frac{a}{b}\right)^2 \left(I_x \otimes \bar{B}_y\right) \tag{A3}$$

$$H_2 = C\left(\bar{A}_x \otimes \bar{A}_y\right) \tag{A4}$$

$$H_3 = E_2\left(I_x \otimes \bar{B}_y\right) + \mu G_{12}\left(\frac{b}{a}\right)^2 \left(\bar{B}_x \otimes I_y\right) \tag{A5}$$

$$H_4 = \bar{B}_x \otimes I_y + \frac{D_2}{D_1}\left(\frac{a}{b}\right)^2 \left(\bar{B}_x \otimes \bar{B}_y\right) + \frac{D_3}{D_1}\left(\frac{a}{b}\right)^4 \left(I_x \otimes \bar{B}_y\right) \tag{A6}$$

$$H_5 = E_1\left(\frac{h}{a}\right)^2 \left(\bar{B}_x \otimes I_y\right) + \upsilon_{12} E_2\left(\frac{h}{b}\right)^2 \left(I_x \otimes \bar{B}_y\right) \tag{A7}$$

$$H_6 = E_2\left(\frac{h}{b}\right)^2 \left(I_x \otimes \bar{B}_y\right) + \upsilon_{21} E_1\left(\frac{h}{a}\right)^2 \left(\bar{B}_x \otimes I_y\right) \tag{A8}$$

$$H_7 = \frac{h^2}{a^2}\left(\overline{A}_x \otimes I_y\right) \tag{A9}$$

$$H_8 = \frac{h^2}{b^2}\left(I_x \otimes \overline{A}_y\right) \tag{A10}$$

**Table 1. The Center deflections of the clamped and simply-supported square plates**

| Methods | Analytical[27] | FEM [28] | FEM [27] | Present |
|---|---|---|---|---|
| Simply supported | 0.940 | 1.028(9.3%) | 0.942(0.3%) | 0.944(0.4%) |
| Clamped | 1.151 | 1.316(14.3%) | 1.170(1.6%) | 1.123(2.4%) |

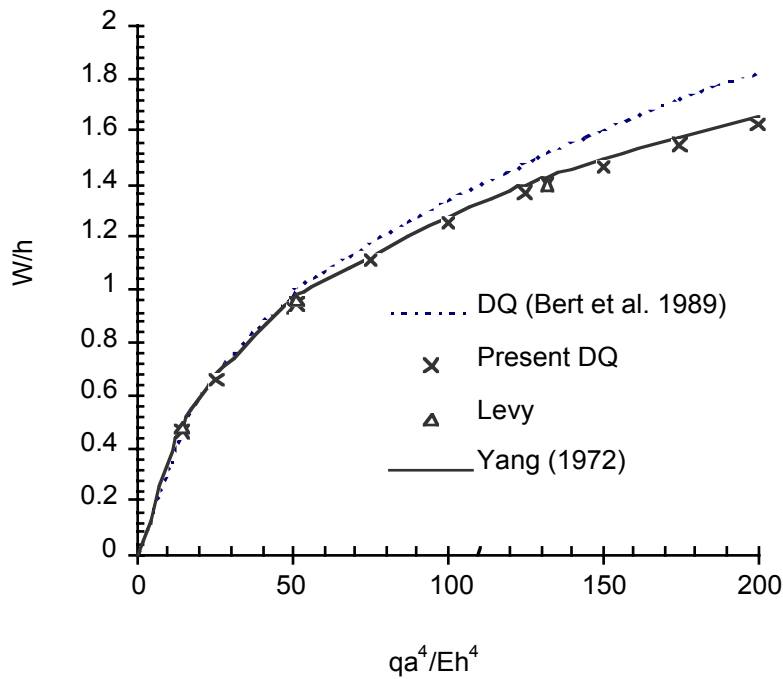

Fig. 1. Central deflections for a simply supported square isotropic plate

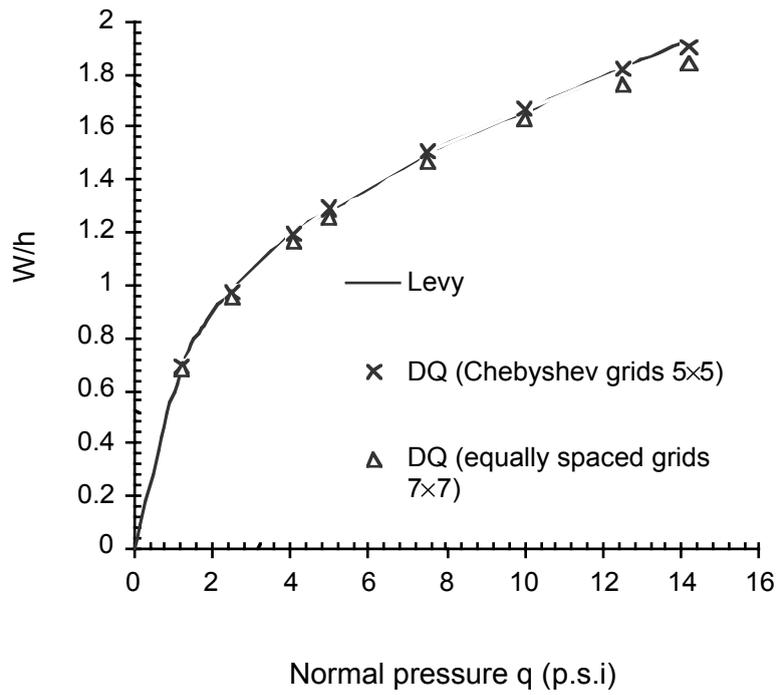

Fig. 2. Comparison of the DQ accuracies of central deflections for a square simply supported plate using the Chebyshev 5×5 and the equally spaced 7×7 grids.

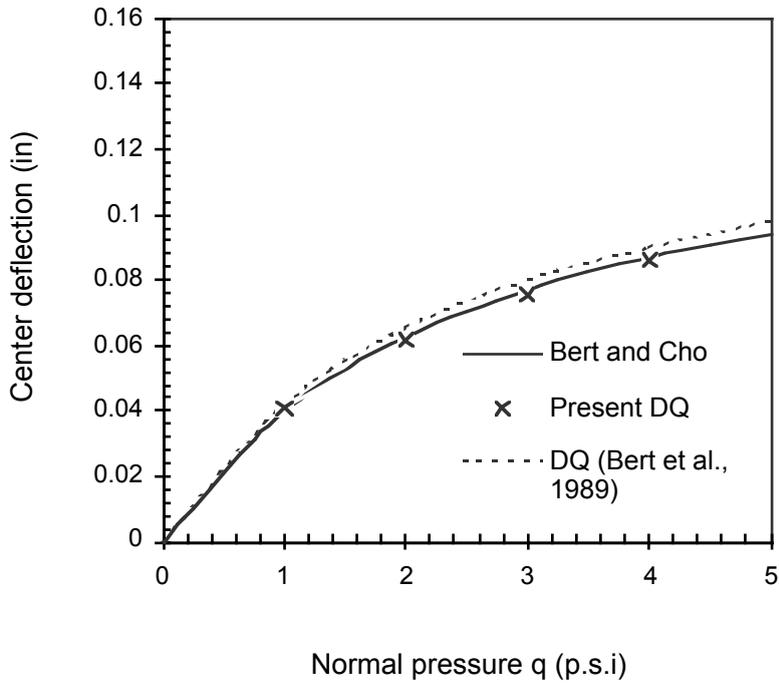

Fig. 3. Central deflections for a clamped square orthotropic plate

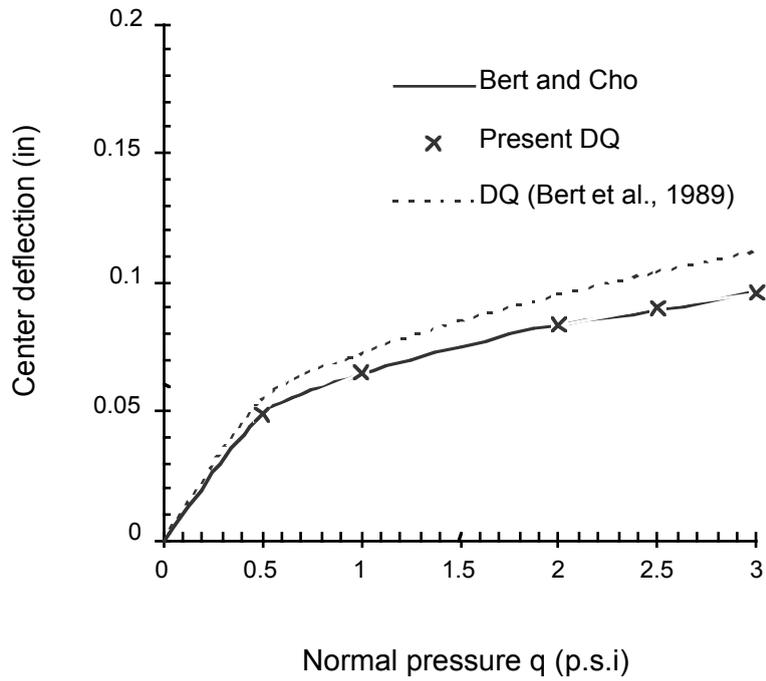

Fig. 4. Central deflections for a simply supported square orthotropic plate.